\documentclass[conference,a4paper]{IEEEtran}

\ifCLASSINFOpdf
\else
\fi
\usepackage[left=1.57cm,right=1.57cm,top=0.95cm,bottom=2.54cm]{geometry}
\usepackage{setspace}
\usepackage{graphicx}
\usepackage{tabularx,booktabs}
\usepackage{dingbat}
\usepackage{diagbox}
\usepackage{multirow} 
\usepackage[hyphens]{url}
\usepackage[hidelinks,breaklinks]{hyperref}
\usepackage{xcolor}
\usepackage{amsfonts, amsmath, amsthm, amssymb}
\usepackage{subcaption}
\hypersetup{breaklinks=true}
\usepackage{tikz}
\usepackage{textcomp}
\usepackage{lipsum}
\usepackage{svg}
\IEEEoverridecommandlockouts
\newcommand\copyrighttext{%
  \footnotesize \textcopyright 2025 IEEE.  Personal use of this material is permitted.  Permission from IEEE must be obtained for all other uses, in any current or future media, including reprinting/republishing this material for advertising or promotional purposes, creating new collective works, for resale or redistribution to servers or lists, or reuse of any copyrighted component of this work in other works.}
\newcommand\copyrightnotice{%
\begin{tikzpicture}[remember picture,overlay]
\node[anchor=south,yshift=10pt] at (current page.south) {\fbox{\parbox{\dimexpr\textwidth-\fboxsep-\fboxrule\relax}{\copyrighttext}}};
\end{tikzpicture}%
}

\usepackage{graphicx}
\graphicspath{{figures/}}
\begin{document}

%
\title{UplinkNet: Practical Commercial 5G Standalone (SA) Uplink Throughput Prediction}

\author{
    \IEEEauthorblockN{Kasidis Arunruangsirilert\IEEEauthorrefmark{1}, Jiro Katto\IEEEauthorrefmark{1}}
    \IEEEauthorblockA{\IEEEauthorrefmark{1}Department of Computer Science and Communications Engineering, Waseda University, Tokyo, Japan
    \\\{kasidis, katto\}@katto.comm.waseda.ac.jp}
}
%

\maketitle
\copyrightnotice
\setstretch{0.95}
\begin{abstract}

While 5G New Radio (NR) networks offer significant uplink throughput improvements, these gains are primarily realized when User Equipment (UE) connects to high-frequency millimeter wave (mmWave) bands. The growing demand for uplink-intensive applications, such as real-time UHD 4K/8K video streaming and Virtual Reality (VR)/Augmented Reality (AR) content, highlights the need for accurate uplink throughput prediction to optimize user Quality of Experience (QoE). In this paper, we introduce \textit{UplinkNet}, a compact neural network designed to predict future uplink throughput using past throughput and RF parameters available through the Android API. With a model size limited to approximately 4,000 parameters, \textit{UplinkNet} is suitable for IoT and low-power devices. The network was trained on real-world drive test data from commercial 5G Standalone (SA) networks in Tokyo, Japan, and Bangkok, Thailand, across various mobility conditions. To ensure practical implementation, the model uses only Android API data and was evaluated on unseen data against other models. Results show that \textit{UplinkNet} achieves an average prediction accuracy of 98.9\% and an RMSE of 5.22 Mbps, outperforming all other models while maintaining a compact size and low computational cost.
\looseness=-1

\end{abstract}



\setstretch{0.93}

%
\IEEEpeerreviewmaketitle

\vspace{-6mm}
\section{Introduction}

In the early smartphone era, most mobile network use cases focused on content consumption, resulting in predominantly downlink (DL) traffic. However, with the rise of social media platforms encouraging real-time content creation and sharing, uplink (UL) throughput demand has surged, particularly for high-bandwidth content such as UHD 4K/8K video, Virtual Reality (VR), and Augmented Reality (AR). The introduction of 5G New Radio (NR) promise to offers substantial improvements in UL throughput over 4G Long-Term Evolution (LTE), promising the peak target of up to 10 Gbps \cite{itu_m2083}. However, such throughput is largely experimental and achievable only in controlled environments. In real-world, the achievable uplink throughput is much lower. For instance, the Qualcomm Snapdragon X75 modem, expected in smartphones from late 2023, supports a maximum UL speed of 3.5 Gbps \cite{qualcomm_2023}, while a recent demonstration by AIS and ZTE achieved 2.12 Gbps \cite{zte_corporation_2023} on the 5G millimeter wave (mmWave) band in a test setup. In contrast, typical commercial deployments in Japan and Thailand, with a 3:1 DL to UL ratio, offer a theoretical peak UL speed of 1 Gbps on 400 MHz of bandwidth. However, most 5G coverage is provided by mid- and low-frequency bands with limited bandwidth, leading to reduced throughput. A recent study \cite{10118777} shows that while the theoretical maximum UL throughput on the mid-frequency band can reach 285.72 Mbps using high-end user equipment (UE), the average real-world throughput can be as low as 9.86 Mbps, with further degradation on devices with a single transmission antenna.

Previous efforts to predict mobile network throughput typically adopt either application-layer approaches, using packet loss and delay \cite{10.1145/2910017.2910608, 10.1145/3300061.3345430, 9488851}, or physical-layer approaches, using radio frequency (RF) parameters \cite{10.1145/3386901.3388911, 8051088, 10147378}. However, many of these approaches require parameters such as Resource Block Allocation (RB) and Transmission Power (Tx Power), which are not accessible without modifying the smartphone, rendering them impractical. Furthermore, existing methods often do not account for different frequency bands and duplex schemes, both of which significantly affect the achievable UL throughput. For example, 5G SA networks use both legacy LTE spectrum and newly allocated spectrum, resulting in the utilization of both Frequency Division Duplexing (FDD) and Time Division Duplexing (TDD) on the same network \cite{ray_2020}. The introduction of Massive MIMO Active Antenna Units (AAUs), with up to 128 antenna elements compared to the maximum of eight in legacy setups, further complicates the effort, particularly due to its impact on beamforming performance and UL throughput \cite{ericsson_2018}.

Given the limited UL throughput in Sub-6 GHz bands, accurate throughput prediction is essential for optimizing user Quality of Experience (QoE) and ensuring reliable connections for mission-critical IoT applications such as autonomous vehicles and industrial automation. Accurate predictions enable effective link adaptation and adaptive data compression \cite{10.1145/3605573.3605588}, especially when the UE enters areas with poor signal quality. In this paper, we propose \textit{UplinkNet}, a compact neural network designed to predict UL throughput in commercial 5G SA networks using only data available via the Android API. The model, constrained to approximately 4,000 parameters to ensure suitability for IoT devices, is trained on real-world data collected from Tokyo, Japan, and Bangkok, Thailand, and evaluated on unseen data from various mobility scenarios, including walking, driving, and riding public transport. This paper is organized as follows: Section II provides background information, Section III presents experimental results and analysis, and Section IV concludes with future directions. \looseness=-1

\section{Background}

\subsection{Network Configuration and Data Collection}

For the UE, Samsung Galaxy S22 Ultra 5G (SC-52C) with Qualcomm Snapdragon X65 5G RF Modem \cite{qualcomm_x65} was used. Due to the rarity of UL-2Tx-capable UE \cite{10118777}, the UL-1Tx UE was chosen. By using Network Signal Guru (NSG), a professional mobile network drive test software, the \textit{UECapabilityInformation} packet was obtained and verified that the UE is compatible with all frequency bands that are being used in both Japan and Thailand. According to the 3GPP standard \cite{3GPP_38-101-1}, due to high path loss at a higher frequency, High Power UE (HPUE) may use higher transmission power on some frequency bands. However, this feature is unsupported on Japanese networks, so it was disabled. Both training and evaluation data were obtained on SoftBank, an MNO in Japan, and Advanced Info Service (AIS), an MNO in Thailand. Both networks provide a meaningful coverage of 5G SA service in their respective area, which allows the data collection to be possible.

By using the NSG test function, the upload stress test was conducted by continuously transmitting the data via HTTP POST protocol to the server. All of the RF parameters were saved to log files. Additionally, the GPS location and mobility speed were also recorded for reference purposes. When applicable, the UE is placed on the side of the vehicle's window to ensure the best signal reception and to eliminate the variable caused by the way the smartphone is being held. Training data consisted of 40 traces with more than 35 hours of real-world drive tests in the capital cities of two countries. Each trace was collected at different times of the day on various types of vehicles and public transportation to cover a variety of mobility speeds. Due to SoftBank's network configuration, UE is usually automatically attached to frequency band n3, so the modem was modified to have band n3 and n28 disabled in some scenarios to collect the data on band n77. On the other hand, AIS barred n28 cells, preventing all UE from connecting altogether. Therefore, all data collected on AIS were from frequency band n41. The testing data was collected using the same procedure. The summary of testing data can be seen in Table \ref{tab:TestingDataSoftBank} and \ref{tab:TestingDataAIS}.


%

\begin{figure}[t!]
\begin{subfigure}{\linewidth}
  \centering\includesvg[width=0.95\linewidth,inkscapelatex=false]{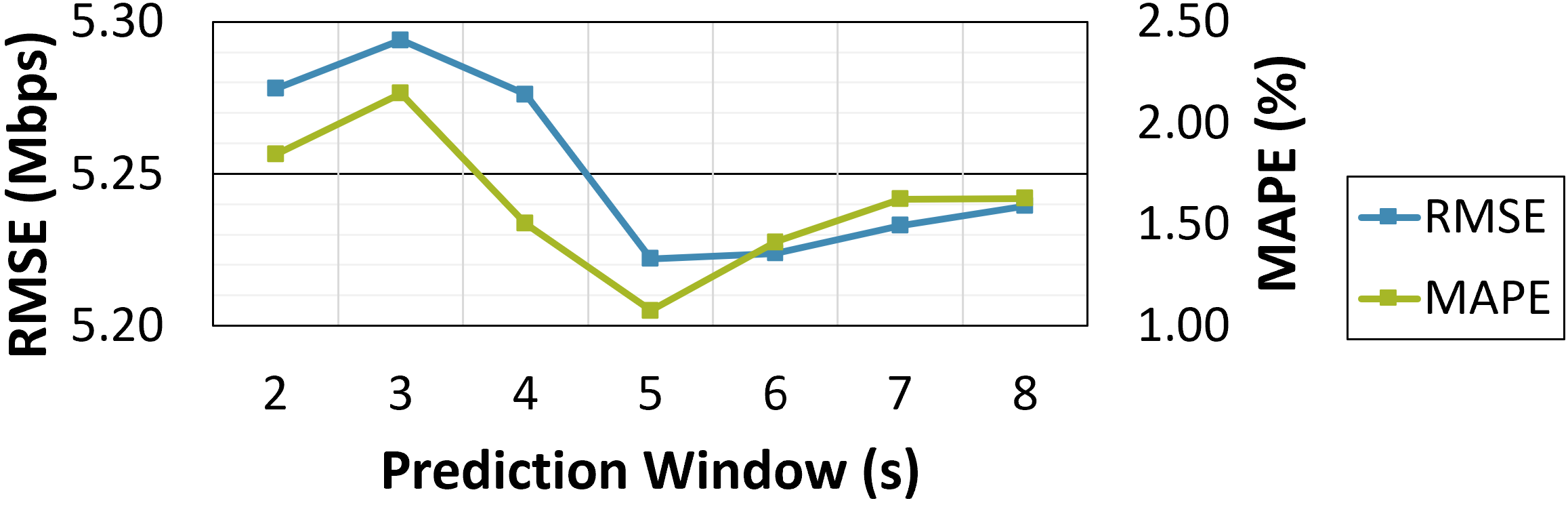}
  \vspace{-1mm}
  \caption{Prediction Window}
  \label{fig:PredictionWindow}
\end{subfigure}\\
\begin{subfigure}{\linewidth}
  \centering\includesvg[width=0.95\linewidth,inkscapelatex=false]{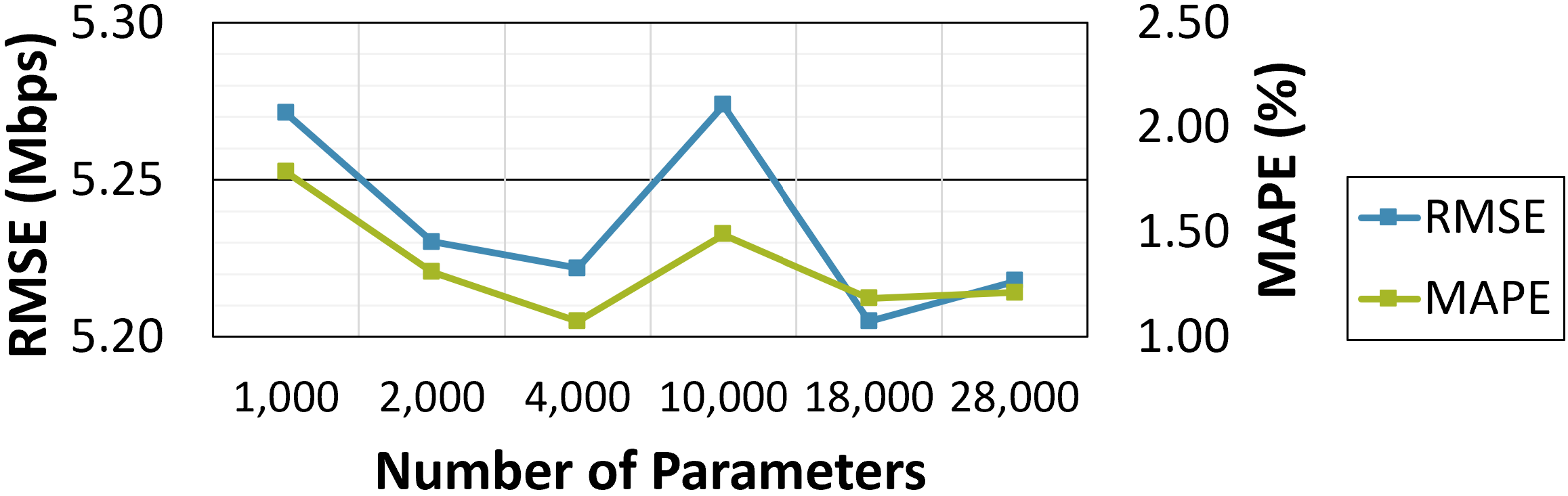}
  \vspace{-0.5mm}
  \caption{Number of Parameters}
  \label{fig:LSTMLayer}
  \vspace{-5mm}
\end{subfigure}\\

\begin{subfigure}[t]{\linewidth}
\centering\includesvg[width=0.25\linewidth,inkscapelatex=false]{PredictionAccuracyLegend.svg}
\end{subfigure}
\vspace{-5mm}
\caption{Prediction Accuracy on Evaluation Data}
\label{fig:ParamsSummary}
\vspace{-4mm}
\end{figure}

\begin{figure}[t!]
\begin{subfigure}{.48\linewidth}
  \centering\includesvg[width=0.95\linewidth,inkscapelatex=false]{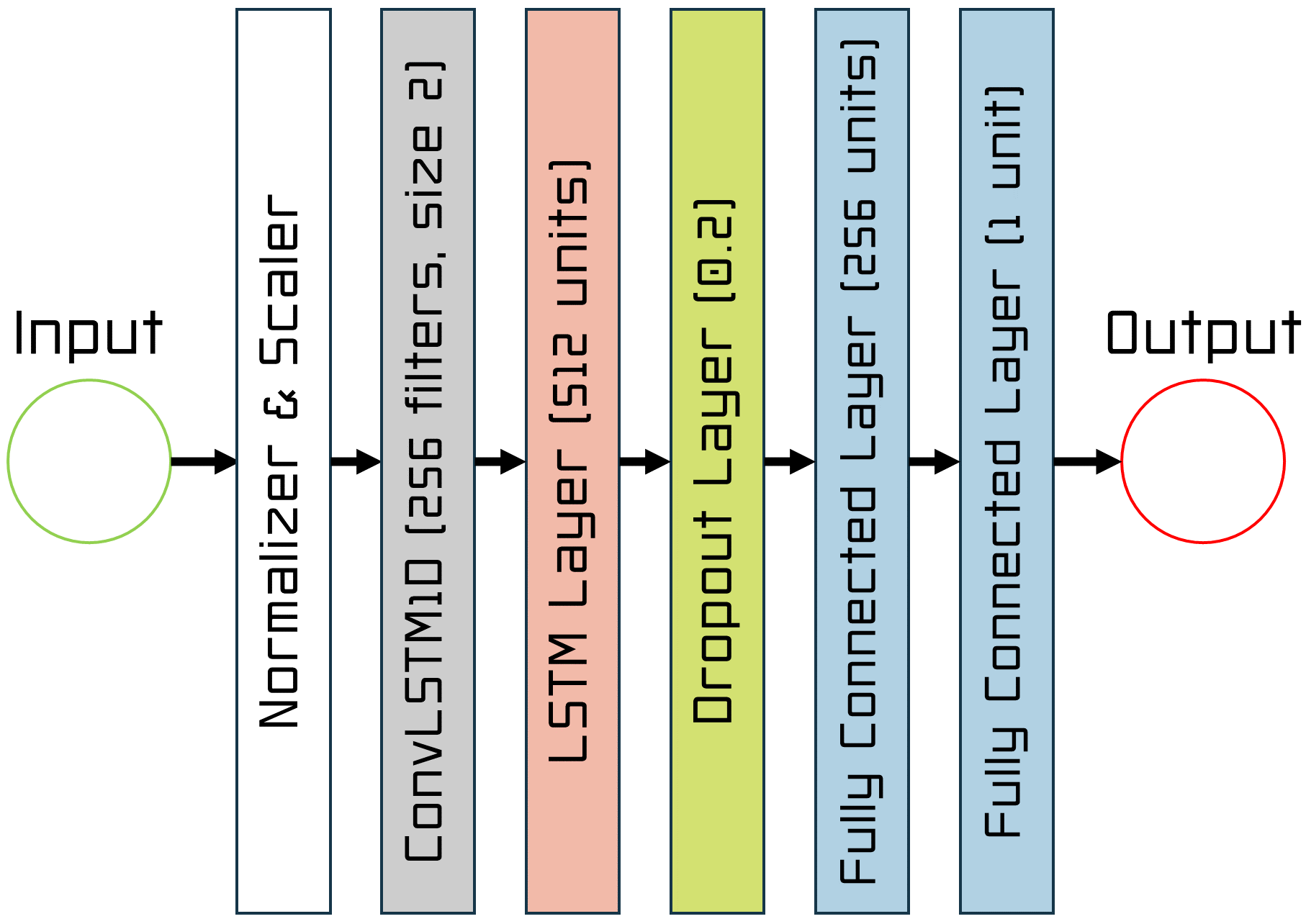}
  \vspace{-0.5mm}
  \caption{UplinkNet (Ours)}
  \label{fig:ConvLSTM}
\end{subfigure}\hfill
\begin{subfigure}{.48\linewidth}
  \centering\includesvg[width=0.95\linewidth,inkscapelatex=false]{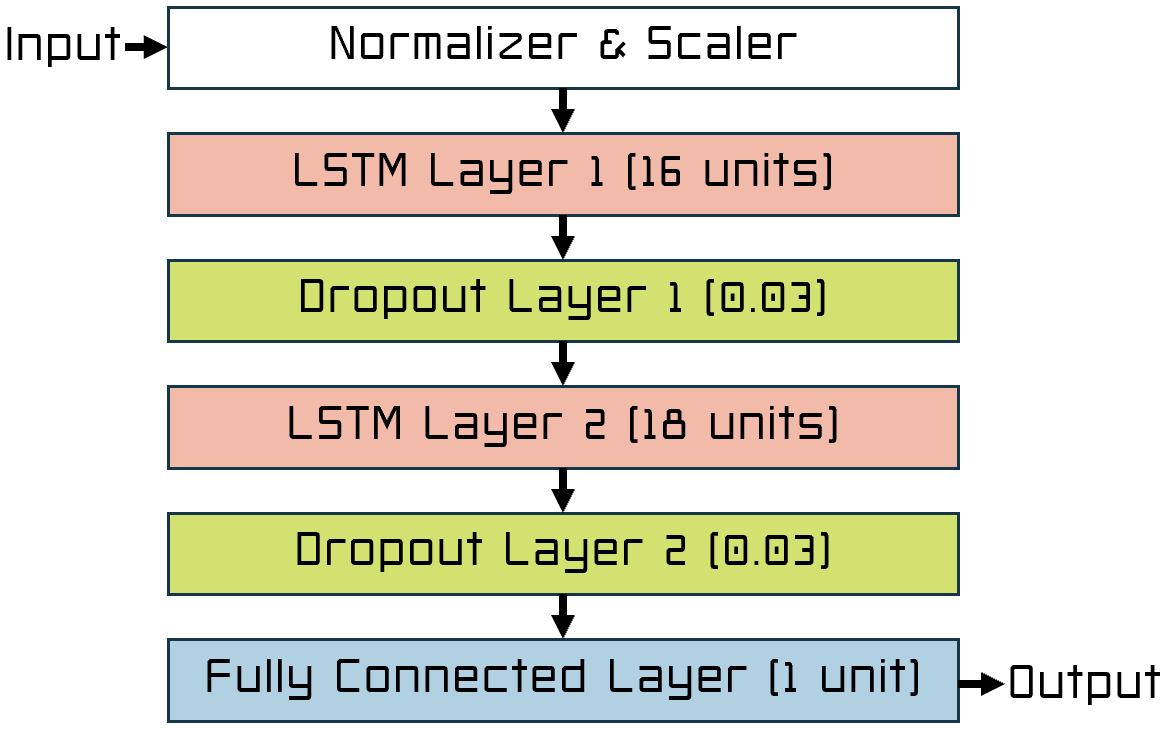}
  \vspace{-0.5mm}
  \caption{LSTM Model}
  \label{fig:LSTM}
  \vspace{1mm}
\end{subfigure}\\
\begin{subfigure}{.48\linewidth}
  \centering\includesvg[width=0.95\linewidth,inkscapelatex=false]{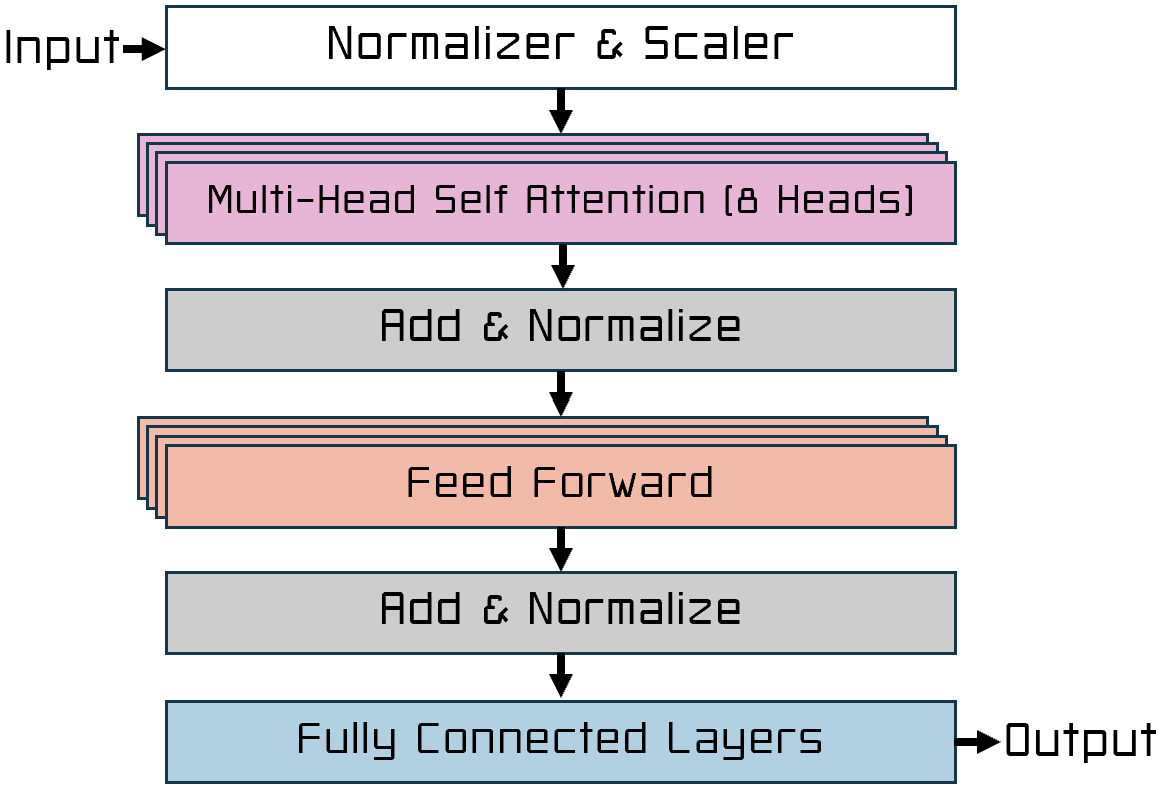}
  \vspace{-0.5mm}
  \caption{Transformer (SURE (4k))}
  \label{fig:Transformer}
\end{subfigure}\hfill
\begin{subfigure}{.48\linewidth}
  \centering\includesvg[width=0.95\linewidth,inkscapelatex=false]{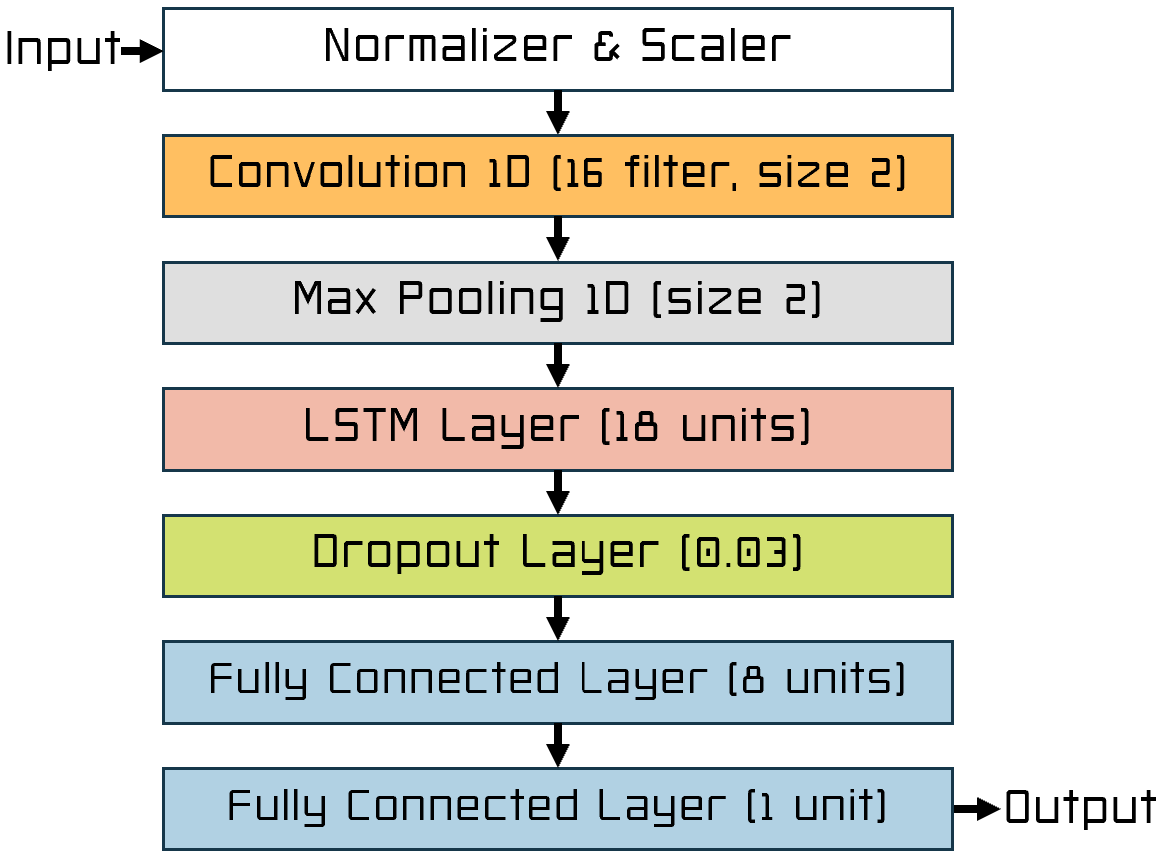}
  \vspace{-0.5mm}
  \caption{CNN-LSTM Model}
  \label{fig:CNNLSTM}
  \vspace{1mm}
\end{subfigure}

\vspace{-1.5mm}
\caption{Summary of Models Architecture}
\vspace{-6mm}
\end{figure}

\begin{table*}[!tbp]
\caption{Summary of Testing Data collected on SoftBank Japan.}
\vspace{-2mm}
\setstretch{0.8}
\centering
\label{tab:TestingDataSoftBank}
\resizebox{18cm}{!}{\begin{tabular}{@{}llllllllllll@{}}
\toprule
Test Scenario & Time & Departure & Arrival & Avg.  & Avg.  & Avg.  & Avg.  & Total& Total Data&Avg. &Frequency Band Percentage\\
 &  &  &  & Speed &  RSRP &  RSRQ &  SINR & Data& Transmitted& Thput.&(0.7G/1.8G/3.4G/3.9G) (\%)\\

&&&&(km/h)& (dBm)&(dBm)&(dB)& Point&(MB)&(Mbps)& \\
\midrule

Keisei SkyAccess Line (Train)*&16:37&Aoto&Narita&55.38&-96.92&-15.60&6.88&3361&2862.8&6.81&74.80 / 20.29 / 4.91 / 0.00                            \\
JR Chuo Line (Rapid) (Train)&18:15&Tokyo&Otsuki&48.04&-90.61&-14.37&11.51&5087&8755.7&13.77&16.83 / 37.88 / 34.20 / 11.09                        \\
JR Chuo Line (Rapid) (Train)&20:56&Takao&Nakano&48.05&-93.94&-14.69&10.24&2694&4084.5&12.13&19.12 / 31.66 / 39.31 / 9.91                    \\
JR Musashino Line (n3 only) (Train)&17:36&Nishi-Funabashi&Nishi-Kokubunji&53.28&-95.61&-17.32&7.59&3857&3738.1&7.75&0.00 / 100.00 / 0.00 / 0.00          \\
JR Musashino Line (n28 only) (Train)&13:15&Nishi-Kokubunji&Nishi-Funabashi&53.94&-97.64&-17.02&5.35&3954&7846.7&15.88&100.00 / 0.00 / 0.00 / 0.00         \\
Driving (Urban Tokyo)&19:47&Nishi-Waseda&Shin-Okubo&19.05&-80.09&-13.57&16.83&484&1264.3&20.90&0.00 / 75.00 / 25.00 / 0.00                          \\
Driving (Urban Tokyo)&20:51&Shin-Okubo&Nishi-Waseda&13.94&-90.55&-15.17&15.13&726&1326.6&14.62&4.68 / 71.21 / 24.10 / 0.00                          \\
Tokyo Sakura Tram&18:05&Waseda&Minowabashi&12.06&-81.27&-13.97&14.52&3572&11091.1&24.84&8.87 / 66.94 / 20.32 / 3.86                 \\
Yurikamome (Metro)&19:53&Shimbashi&Toyosu&27.69&-81.47&-15.66&9.58&1727&6047.1&28.01&28.84 / 45.80 / 8.05 / 17.31                           \\
Yurikamome (Metro)&20:23&Toyosu&Shimbashi&24.41&-75.71&-14.79&10.01&2166&8994.6&33.22&14.17 / 63.71 / 9.19 / 12.93                          \\
Walking&21:08&Shiodome&Shimbashi&3.29&-82.29&-15.27&14.48&1590&7783.6&39.16&44.78 / 22.83 / 2.20 / 30.19                            \\
Tohoku Shinkansen (High-Speed Rail) &10:00&Tokyo&Utsunomiya&136.88&-89.18&-15.43&9.79&2618&4501.6&13.76&33.73 / 39.80 / 16.27 / 10.20                            \\

\bottomrule
\end{tabular}}
\vspace{-3mm}
\end{table*}

\begin{table*}[!tbp]
\caption{Summary of Testing Data collected on AIS Thailand.}
\vspace{-2mm}
\setstretch{0.8}
\centering
\label{tab:TestingDataAIS}
\resizebox{14cm}{!}{\begin{tabular}{@{}lllllllllll@{}}
\toprule
Test Scenario & Time & Departure & Arrival & Avg.  & Avg.  & Avg.  & Avg.  & Total& Total Data&Avg. \\
 &  &  &  & Speed &  RSRP &  RSRQ &  SINR & Data& Transmitted& Thput.\\

&&&&(km/h)& (dBm)&(dBm)&(dB)& Point&(MB)&(Mbps)\\
\midrule
SRT Airport Rail Link (Train) & 20:00 & Phaya Thai & Suvarnabhumi & 61.39 &-94.75 & -14.46 & 9.75 & 1599 & 3038.3 & 15.20\\
Driving (Suburb) & 8:52 &Bang Bo & Minburi & 56.85 & -89.23 & -14.12 & 19.12 & 3324 & 10963.7 & 26.39\\
Driving (Urban Bangkok)* & 14:16 & Lat Krabang & Don Mueang & 70.53 & -83.52 & -14.36 & 15.10 & 2166 & 7958.2 & 29.39\\

\bottomrule
\end{tabular}}
\vspace{-4mm}
\end{table*}

\begin{figure*}[t!]
\begin{subfigure}[t]{.32\linewidth}
  \centering\includesvg[width=0.93\linewidth,inkscapelatex=false]{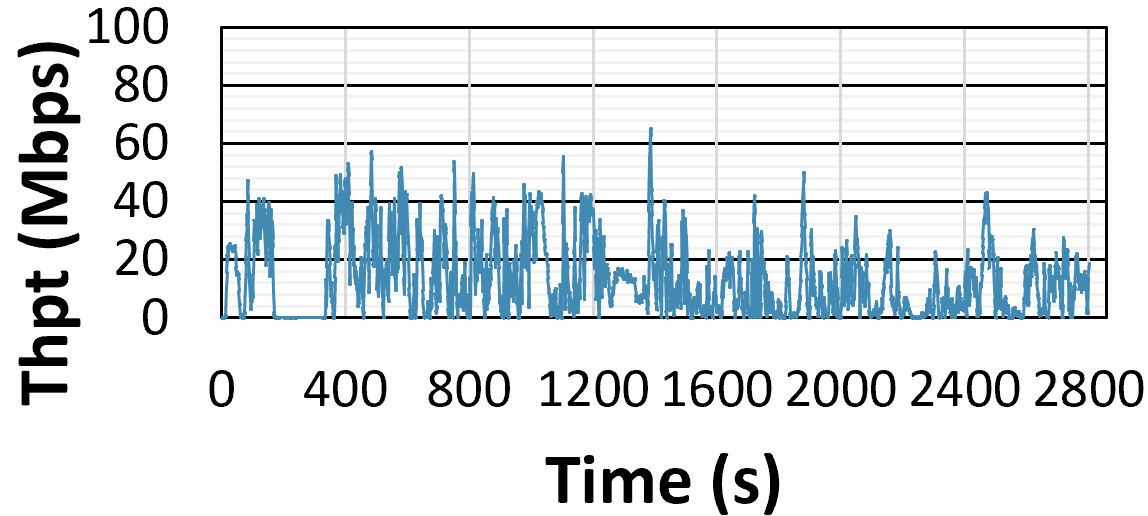}
  \vspace{-1mm}
  \caption{Shinkansen (High-Speed Rail)}
  \label{fig:TrainThpt}
\end{subfigure}\hfill
\begin{subfigure}[t]{.32\linewidth}
  \centering\includesvg[width=0.93\linewidth,inkscapelatex=false]{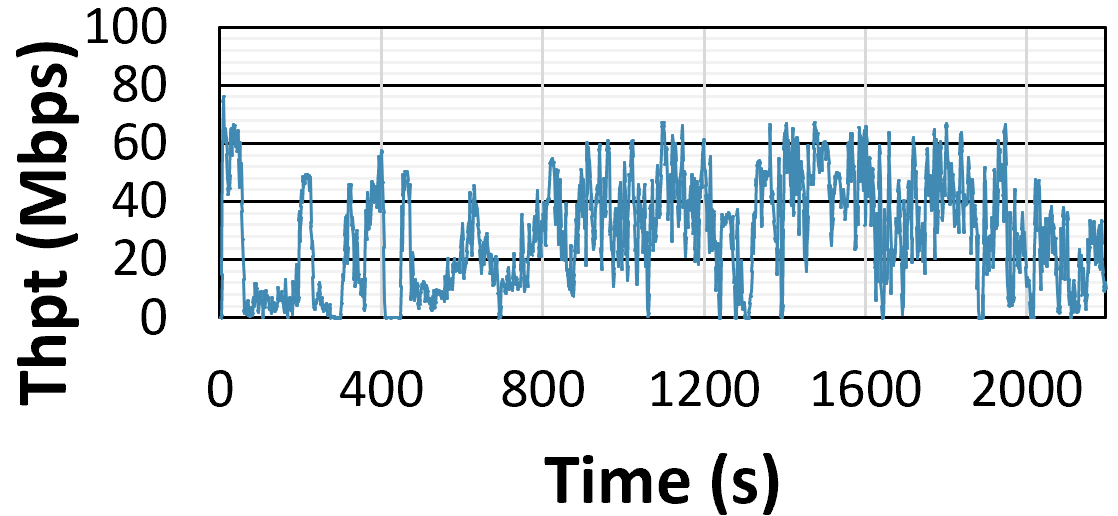}
  \vspace{-1mm}
  \caption{Driving (Urban Bangkok)}
  \label{fig:DrivingThpt}
\end{subfigure}\hfill
\begin{subfigure}[t]{.32\linewidth}
  \centering\includesvg[width=0.93\linewidth,inkscapelatex=false]{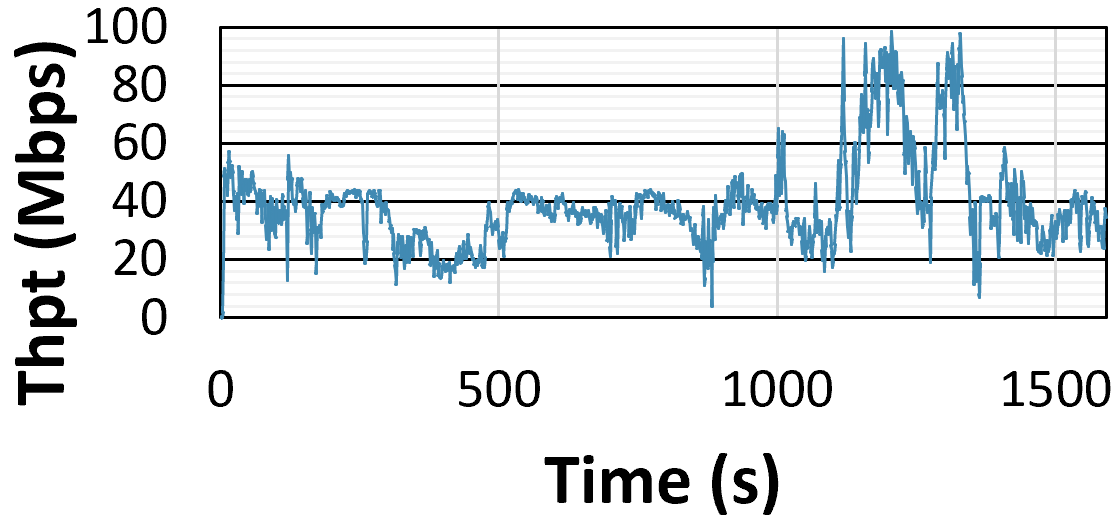}
  \vspace{-1mm}
  \caption{Walking in Shimbashi}
  \label{fig:WalkingThpt}
\end{subfigure}

\setlength{\belowcaptionskip}{-18pt}
\vspace{-1mm}
\caption{5G Uplink Throughput characteristics in some of the test scenarios.}
\vspace{-0.5mm}
\end{figure*}

\vspace{-2mm}

\subsection{Neural Networks and Input Parameters}

\begin{figure}[t!]
  \centering\includesvg[width=0.93\linewidth,inkscapelatex=false]{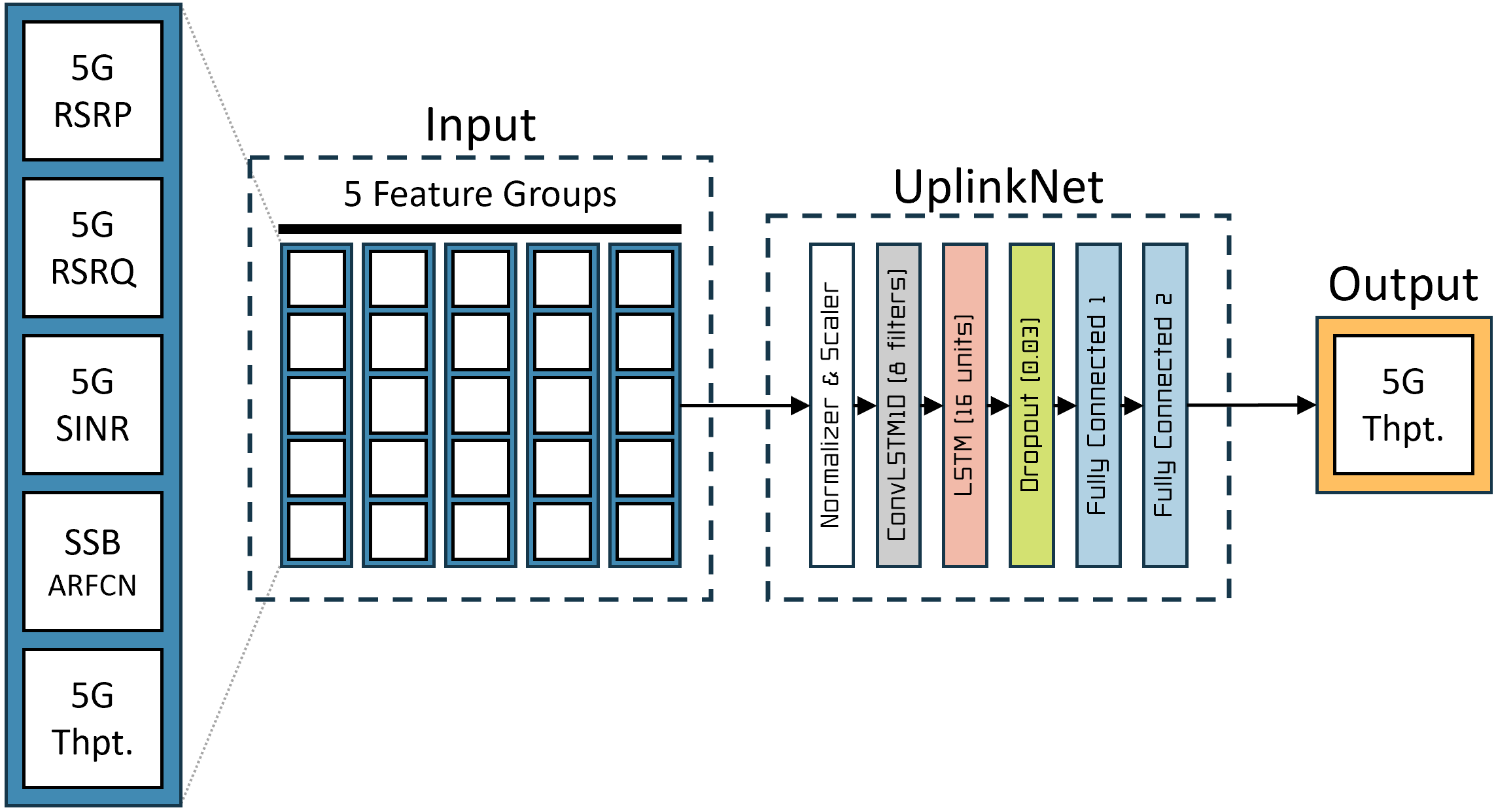}
  \setlength{\belowcaptionskip}{-5pt}
  \vspace{-1mm}
  \caption{Overview of UplinkNet}
  
  \label{fig:System}
\vspace{-4mm}

\end{figure}

ConvLSTM layer \cite{10.5555/2969239.2969329} was used as the input layer of the UplinkNet, then the LSTM layer was used as the hidden layer. Finally, the output layer consisted of a fully connected layer (see Figure \ref{fig:ConvLSTM}). The model structure was configured in a way that the number of parameters is around 4,000 parameters (see Figure \ref{fig:LSTM} and \ref{fig:CNNLSTM}). For comparison, similar networks with LSTM and CNN-LSTM as the input layers were also implemented targeting the same number of parameters. ConvLSTM layer, in Figure \ref{fig:ConvLSTM}, has an advantage over CNN-LSTM, in Figure \ref{fig:CNNLSTM}, when it comes to capturing spatio-temporal relationships as it implements the convolutions operations at gates in LSTM layer directly instead of sequential operation in CNN-LSTM where data is pre-processed using CNN before feeding into LSTM. On the other hand, the conventional LSTM layer (as in Figure \ref{fig:LSTM}) is a one-dimensional operation, so it can only exploit temporal relationships in the input data, but due to this, the computing complexity is lower than CNN-LSTM and ConvLSTM counterparts. 

To compare our proposed model to other works, Best Fixed LSTM (BF-LSTM) based on PERCEIVE \cite{10.1145/3386901.3388911} and Transformer-based model (Self-Attention) similar to SURE \cite{10147378} were also implemented (see Figure \ref{fig:Transformer}). From now on, these implementations will be referred to as PERCEIVE and SURE, respectively. Models from other works were configured both with the same model structure as the original work and with the number of parameters constrained to 4,000 parameters by reducing model width proportionally. The cut-down models will be referred to as PERCEIVE (4k) and SURE (4k). Since the network from the other work accepts different parameters as the input and output Transport Block Size (TBS) instead of throughput in Mbps, it has been modified to match our evaluation scenarios.


As for the prediction time window, a preliminary study was conducted. With the data sample interval of one point per second, it was found that a time window of five seconds yields the best result (see Figure \ref{fig:PredictionWindow}). Furthermore, the prediction accuracy at a different number of parameters was also evaluated and can be seen in Figure \ref{fig:LSTMLayer}. The optimal model configuration from the preliminary study was used. In all cases, the seed for number randomization in Numpy, TensorFlow, and Python were set to 8888, so that the results are reproducible. All models were trained for 200 epochs, except for the transformer model, which was trained for 300 epochs. The learning rate of 0.0001 was used for all models. Early stopping and dropout layers were used to prevent overfitting. We used a dropout rate of 0.03 for all models, except the one that based on PERCEIVE, where the dropout rate of 0.50 used in the original work is maintained. Two of the evaluation data (marked with asterisks in Table \ref{tab:TestingDataSoftBank} and \ref{tab:TestingDataAIS}) were used as the testing data during training, and then the best model was kept for evaluation. Finally, the Adam optimizer was used for the training.
\begin{figure}[t!]
\centering\includesvg[width=0.93\linewidth,inkscapelatex=false]{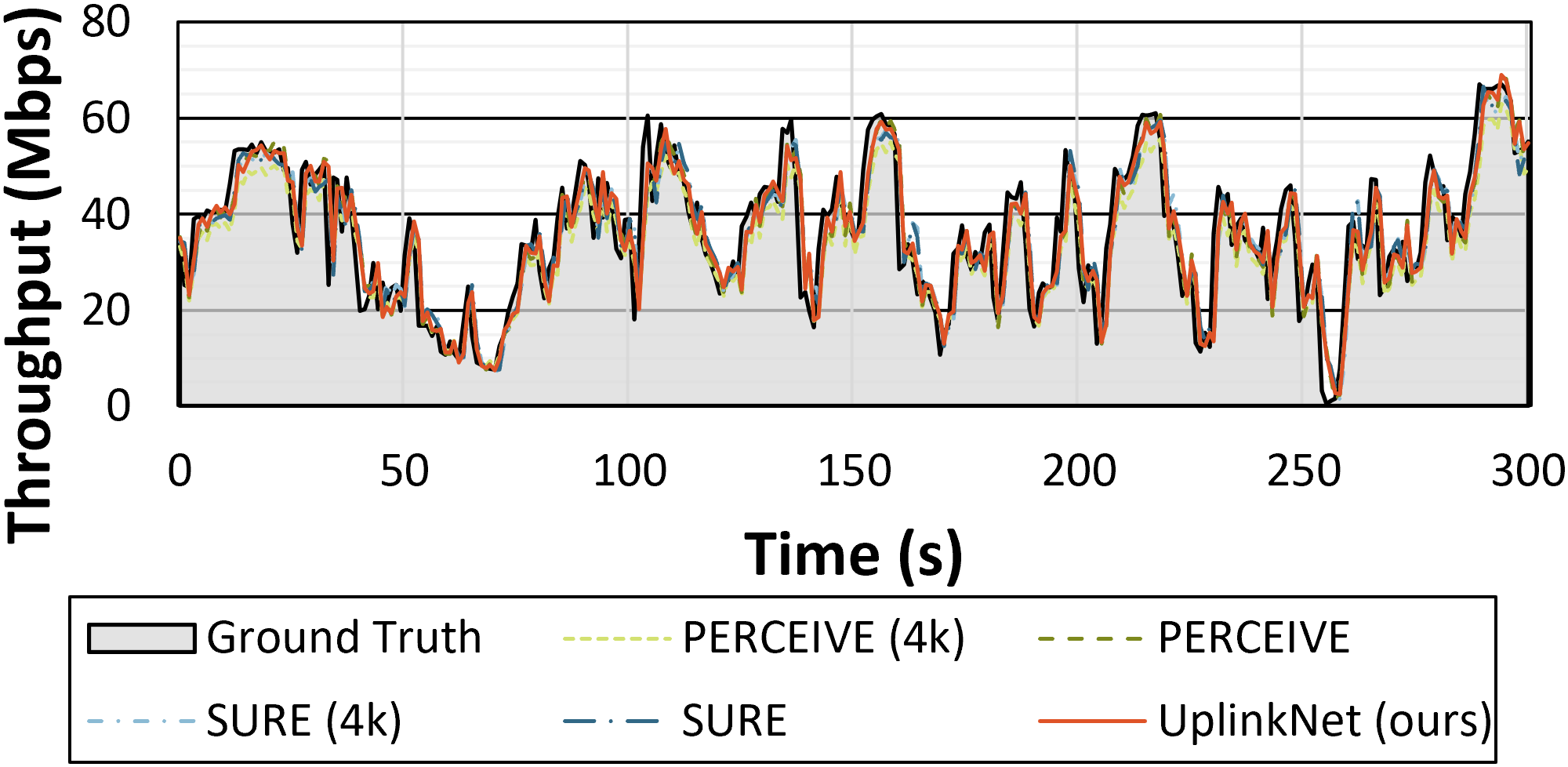}
  \setlength{\belowcaptionskip}{-14pt}
  \vspace{-1.5mm}
  \caption{Predicted uplink throughput from our model compared to others when predicting Driving (Urban Bangkok).}
  \label{fig:GraphPred}
  \vspace{-1mm}
\end{figure}

\begin{figure*}[t!]
\centering
\begin{subfigure}{\linewidth}
\centering\includesvg[width=0.987\linewidth,inkscapelatex=false]{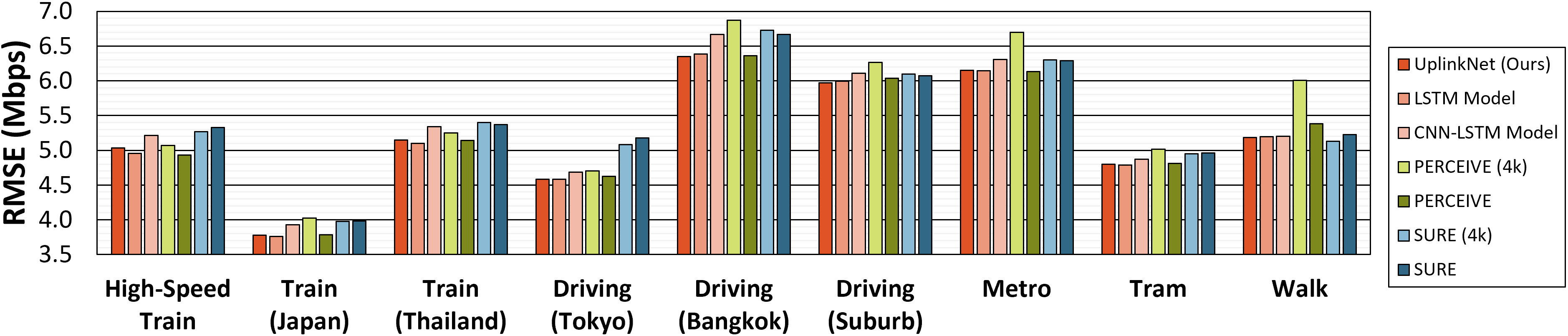}
  \captionsetup{justification=centering}
  \caption{RMSE comparison between each UplinkNet with different input layer, full and parameter-limited PERCEIVE and SURE across nine test scenarios. }
  \label{fig:Results}
\end{subfigure}\\
\begin{subfigure}{.48\textwidth}
  \centering\includesvg[width=1\linewidth,inkscapelatex=false]{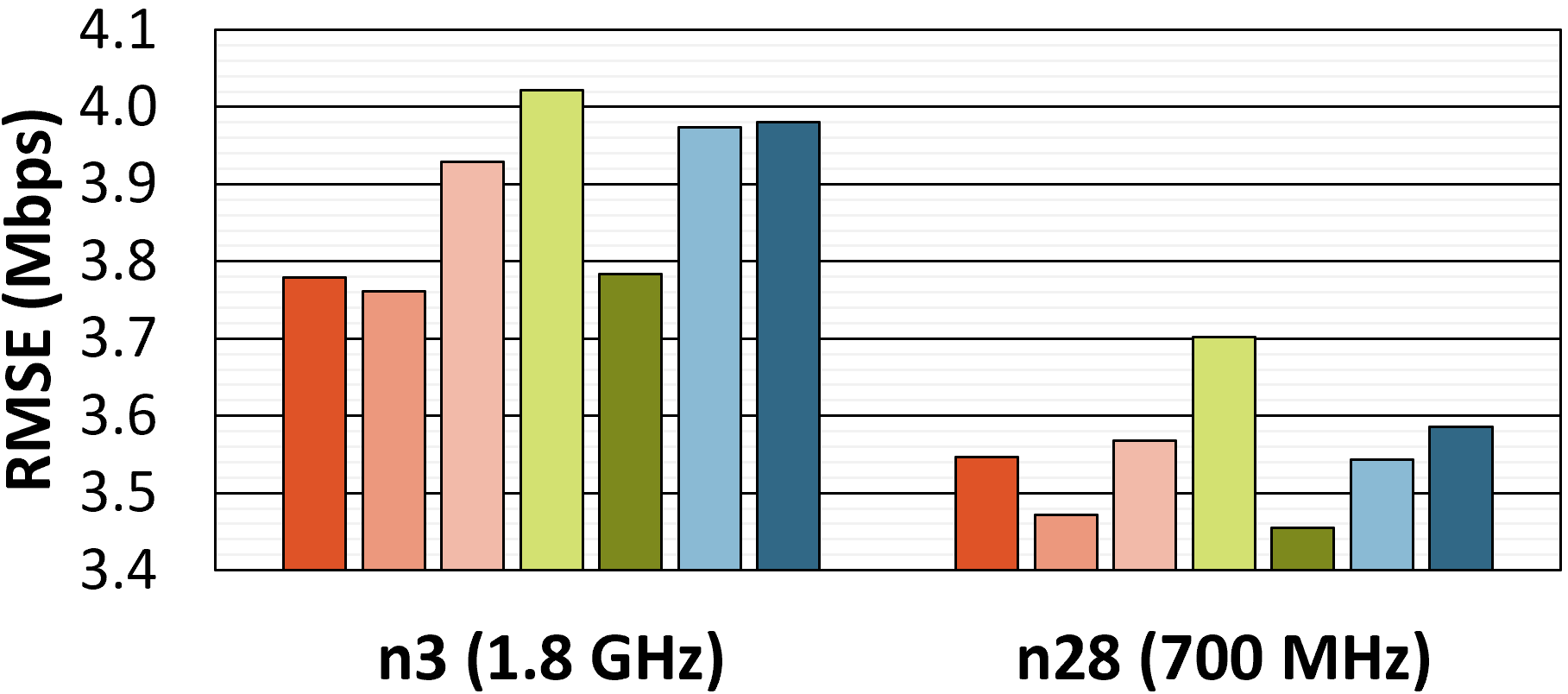}
  \caption{RMSE when UE only support a specific frequency band.}
  \label{fig:BandLimited}
\end{subfigure}%
\hspace{0.02\textwidth}
\begin{subfigure}{.48\textwidth}
  \centering\includesvg[width=1\linewidth,inkscapelatex=false]{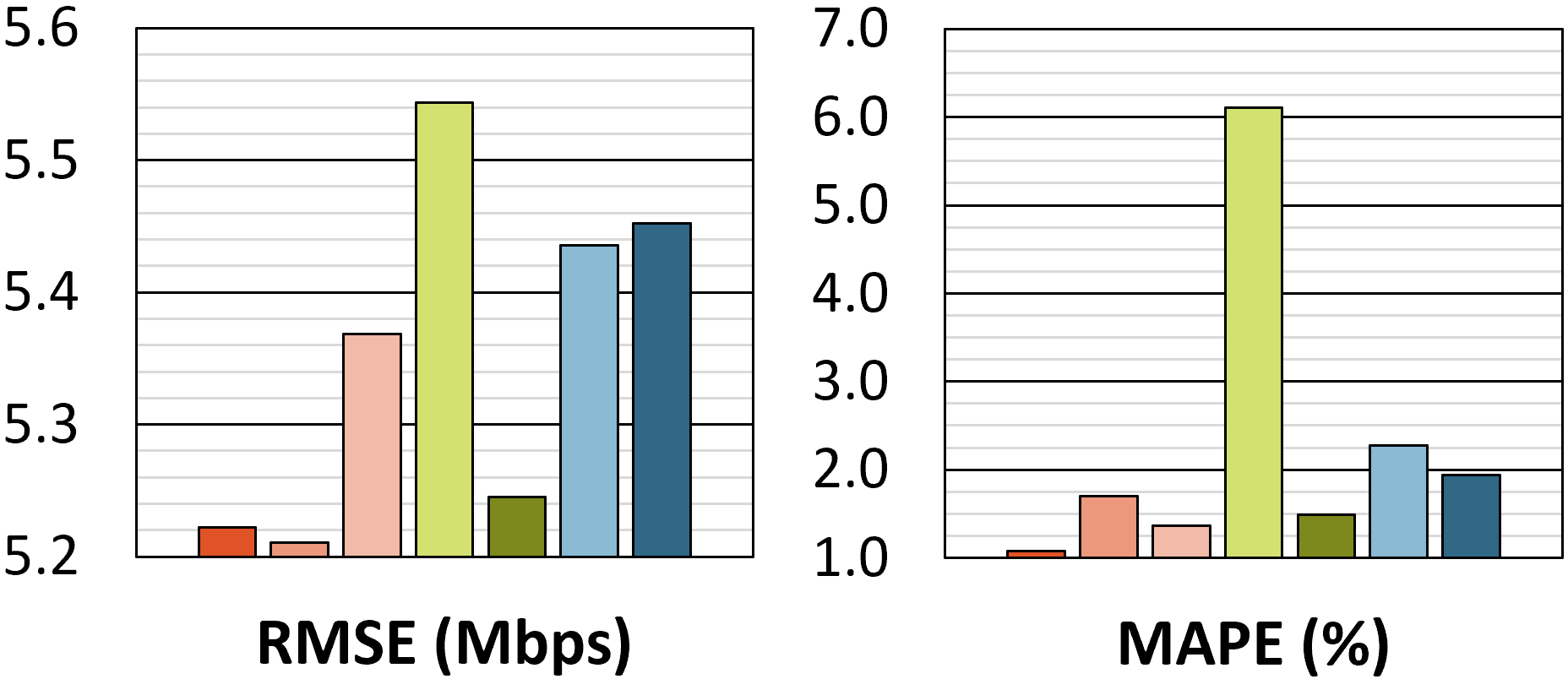}
  \caption{Average Predicted RMSE and MAPE across nine test scenarios.}
  \label{fig:Average}
\end{subfigure}

\vspace{-1mm}

\caption{Evaluation Results}
\vspace{-6mm}

\end{figure*}

\begin{figure}[t!]
\centering
\begin{subfigure}{.24\textwidth}
\centering\includesvg[width=0.987\linewidth,inkscapelatex=false]{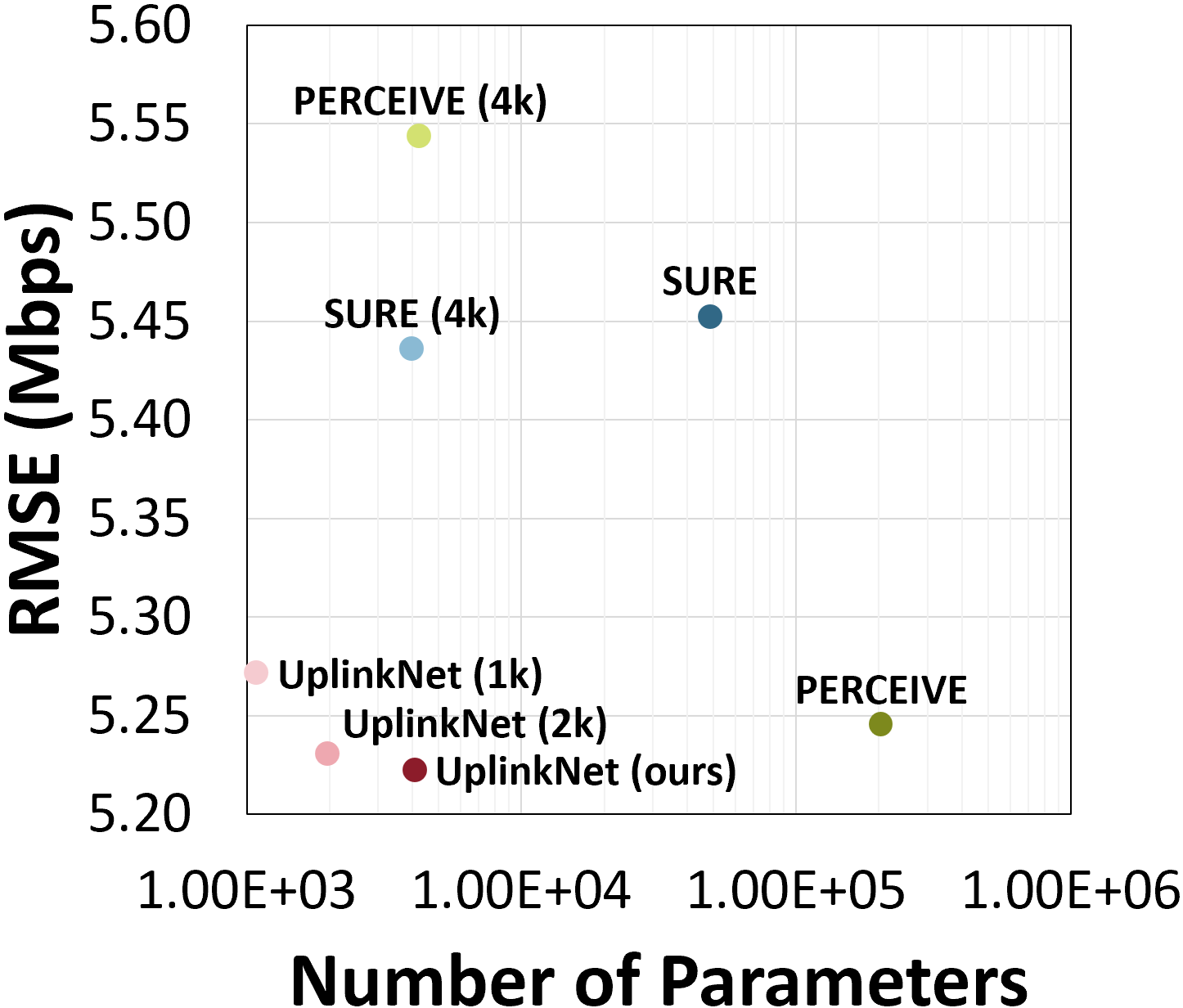}
  \caption{RMSE vs Num of Params}
  \label{fig:paramsRMSE}
\end{subfigure}\hfill
\begin{subfigure}{.24\textwidth}
\centering\includesvg[width=0.987\linewidth,inkscapelatex=false]{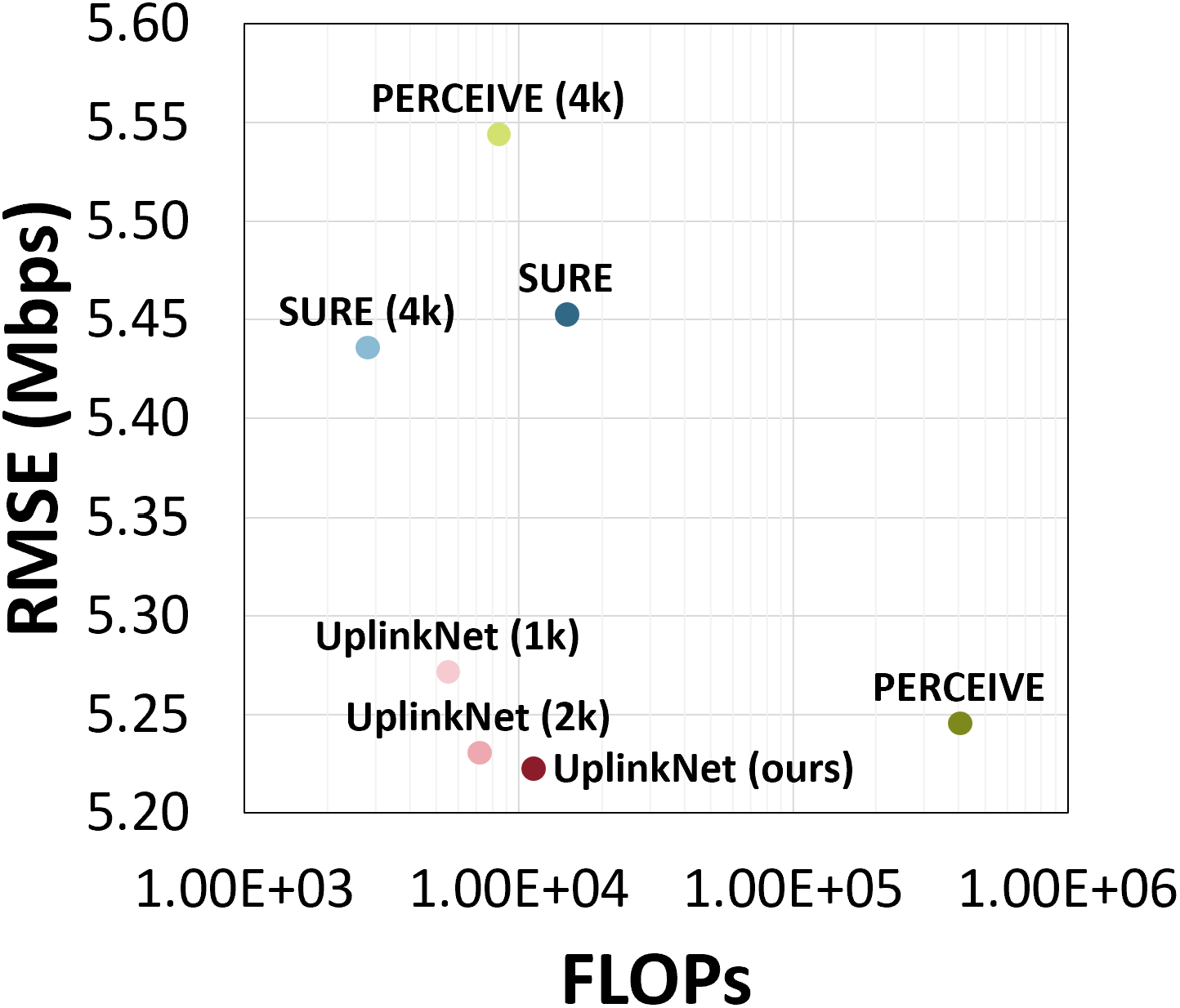}
  \caption{RMSE vs FLOPs}
  \label{fig:FLOPRMSE}
\end{subfigure}
\setlength{\belowcaptionskip}{-20pt}
\vspace{-5mm}
\caption{Complexity comparison between UplinkNet and others}

\end{figure}

While data of all RF parameters with high sampling intervals from the modem chipset are available when using the professional network drive test tools on a modified smartphone, application developers rely on Android API to get information about the RF conditions, which only outputs limited data with very limited update intervals. Even though the correlation has been found between low-level RF parameters such as Tx Power, Resource Block (RB), and uplink throughput \cite{10118777,10.1145/3386901.3388911,10147378,9495144}, these information are not available to typical users. Since the real-world implementation is highly desirable, the data was collected with the sample interval of one second per data point to match the update interval of signal parameters via Android API, and only the parameters that are available via the API; CSI-RSRP, CSI-RSRQ, CSI-SINR, SSB ARFCN (Frequency), and past throughput were used. Despite lacking information about the cell load information, which can easily be derived from RB allocation, the study shows that RSRQ contains the information about the cell load and may be used to predict such information \cite{8506494}. To keep the data collection simple, the data was collected using NSG, and not directly from Android API, but since this set of parameters matches the output information of the API, it will now be referred to as \textit{Android API Data}. Finally, the overall system architecture can be seen in Figure \ref{fig:System}.


\vspace{-1.0mm}
\subsection{Evaluation Metrics and Data}

Two metrics, RMSE and MAPE, will be used to evaluate the prediction accuracy of the models. While Root Mean Square Error (RMSE) can be used to paint the picture of instantaneous throughput prediction accuracy directly, applying mean absolute percentage error (MAPE) to the predicted throughput data directly will be problematic as 5G uplink throughput often hits zero during handover in the area with weak signal due to failure of RACH procedure. Therefore, MAPE is applied to the total amount of transferred data instead. All of the prediction output data points with negative values will be normalized to zero as it's impossible to have a negative throughput.

As seen in Table \ref{tab:TestingDataSoftBank} and \ref{tab:TestingDataAIS}, 15 traces will be used for model performance evaluation. This includes 12 traces from Japan and 3 traces from Thailand. In some of the test cases, multiple traces were used for evaluation to increase the evaluation accuracy. The weighted average will be taken for Keisei SkyAccess Line and JR Chuo Line and referred to as Train (Japan). Data from JR Musashino Line with Frequency Band Lock will be evaluated separately and will not be included in the total average as this is to simulate the case where low-end UE or UE belonging to visiting tourists are used, which may not support all of the 5G frequency bands. Similarly, the weighted average will also be taken for two driving traces in the Tokyo urban area and two Yurikamome (Metro) traces. The average prediction error will contain the average from nine categories excluding two JR Musashino Line traces.


\vspace{-2mm}

\section{Results and Analysis}

Comparing UplinkNet to the others, our model delivers superior prediction accuracy when performing throughput prediction using only the data from Android API. While all of the methods displays slight prediction delay when compared to the ground truth, the predicted throughput trace in Figure \ref{fig:GraphPred} shows that our model is resilient to extreme fluctuation of throughput and most accurately follows the actual uplink throughput. Additionally, it is observed that PERCEIVE (4k) tends to slightly underutilize the available link capacity, while SURE and SURE (4k) tend to overshoot the actual available capacity slightly, especially during peaks and spikes. Interestingly, when there is a sudden drop in throughput, all of the models overshoot the actual throughput by roughly the same amount.

\vspace{-0.5pt}

When looking at the prediction accuracy of each test scenario, UplinkNet outperformed both PERCEIVE and SURE with 204,201 and 48,942 parameters, despite having only 4,113 parameters, reaching RMSE between 2.67 Mbps and 6.35 Mbps with an average RMSE of 5.22 Mbps across nine test scenarios (see Figure \ref{fig:Results} and \ref{fig:Average}). This is compared to the average RMSE of 5.24 Mbps and 5.45 Mbps achieved by the full PERCEIVE and SURE models, respectively. If reduced complexity is preferred, the number of parameters of UplinkNet can be reduced to 1,969 and 1,089 parameters, which still outperformed PERCEIVE (4k) and SURE (4k) by a significant margin, yielding RMSE of 5.23 Mbps and 5.27 Mbps (see Figure \ref{fig:paramsRMSE} and \ref{fig:FLOPRMSE}), respectively. When looking at the prediction of data transfer, we have found that UplinkNet reached the MAPE between 0.07\% and 2.67\% with an average of 1.07\%, also outperformed PERCEIVE and SURE, which achieved the average MAPE of 1.49\% and 1.93\%, respectively. We have found that when reducing the number of parameters of PERCEIVE to 4,000 parameters, the prediction accuracy of the model dropped significantly, while the SURE model maintained a similar performance to the original model. Replacing the input layer of our model with an LSTM layer can improve the instantaneous prediction accuracy by 0.01 Mbps, but the prediction of total data transfer took a hit and degraded by 0.62\%. Using CNN-LSTM as the input layer yields the worst overall result. 

Finally, when considering the case when UE doesn't support all of the frequency bands, we have found that all models displayed degraded performance. Nevertheless, UplinkNet performed well when limited to the frequency band n3 (see Figure \ref{fig:BandLimited}), leading the pack with the RMSE of 3.78 Mbps, but falling behind other models slightly when limited to the frequency band n28. PERCEIVE model best handled this scenario, but the model size is 51 times larger than ours.


\vspace{-1mm}
\section{Conclusions and Future Work}

In this paper, we propose a UplinkNet, a compact neural network model, to predict the throughput using only the RF parameters accessible via Android API to ensure that the implementation is feasible for real-world applications. More than 35 hours of real-world RF parameters on a commercial 5G Standalone (SA) network were obtained using a network drive test tool from various types of transportation in the capital city of two countries and then used to train the model. The model was then evaluated using another set of real-world data on various kinds of transportation such as high-speed trains, trains, trams, metros, cars, and walking against various different types of models used in earlier literature including PERCEIVE and SURE. Many different types of input layers were also experimented with, yielding slightly different results. \looseness=-1

The results show that our model can accurately predict the uplink throughput on a commercial 5G SA network with extreme throughput fluctuation, dropout, and blind spots, when the input data is limited to what is offered by Android API, achieving an average RMSE of 5.22 Mbps when considering the instantaneous throughput and 98.9\% accuracy when considering the total amount of data transferred, achieving high prediction accuracy and outperforming all other models across nine test scenarios, while maintaining a small and compact model size of around 4,000 parameters with the possibility of further parameter reduction if preferred. Therefore, it's highly suitable for implementation in smartphone applications for a variety of use cases on the 5G mobile network including real-time video transmission, self-driving vehicle, and large file transfer. Furthermore, due to its compact size, it's also highly suitable for implementation in low-powered IoT devices operating on a 5G network.

As for future work, the model may be improved to incorporate supports for throughput prediction with newer 5G uplink throughput enhancement features, such as Uplink MIMO and Uplink Carrier-Aggregation, enabled and active on both the network and the UE side as well adding the support of the upcoming 5G Frequency Range 2 (mmWave) SA network, which has been actively tested by MNOs and RAN manufacturers around the world, recently.

\vspace{-1mm}
\section*{Acknowledgement}

This paper is supported by the Ministry of Internal Affairs and Communications's project for efficient frequency utilization toward wireless IP multicasting. Additionally, the authors would like to express their gratitude to \textbf{PEI Xiaohong} of \textit{Qtrun Technologies} for providing \textit{Network Signal Guru (NSG)} and \textit{AirScreen}, the cellular network drive test software used for result collection and analysis in this research.




%
\setstretch{1}
\Urlmuskip=0mu plus 1mu\relax
\bibliographystyle{IEEEtran}
\bibliography{b_reference}

@inproceedings{10.1145/3605573.3605588,
author = {Lin, Fangzheng and Arunruangsirilert, Kasidis and Sun, Heming and Katto, Jiro},
title = {Recoil: Parallel RANS Decoding with Decoder-Adaptive Scalability},
year = {2023},
isbn = {9798400708435},
publisher = {Association for Computing Machinery},
address = {New York, NY, USA},
url = {https://doi.org/10.1145/3605573.3605588},
doi = {10.1145/3605573.3605588},
booktitle = {Proceedings of the 52nd International Conference on Parallel Processing},
pages = {31–40},
numpages = {10},
location = {Salt Lake City, UT, USA},
series = {ICPP '23}
}

@misc{ericsson_2018, title={Whitepaper on antenna system for 5G networks}, url={https://www.ericsson.com/en/reports-and-papers/white-papers/advanced-antenna-systems-for-5g-networks}, journal={www.ericsson.com}, author={Ericsson}, year={2018}, month={Nov} }

@INPROCEEDINGS{8506494,
  author={Raida, Vaclav and Lerch, Martin and Svoboda, Philipp and Rupp, Markus},
  booktitle={2018 Network Traffic Measurement and Analysis Conference (TMA)}, 
  title={Deriving Cell Load from RSRQ Measurements}, 
  year={2018},
  volume={},
  number={},
  pages={1-6},
  doi={10.23919/TMA.2018.8506494}}

@ARTICLE{9495144,
  author={Minovski, Dimitar and Ögren, Niclas and Mitra, Karan and Åhlund, Christer},
  journal={IEEE Transactions on Mobile Computing}, 
  title={Throughput Prediction Using Machine Learning in LTE and 5G Networks}, 
  year={2023},
  volume={22},
  number={3},
  pages={1825-1840},
  doi={10.1109/TMC.2021.3099397}}

@ARTICLE{10147378,
  author={Jung, Jewon and Lee, Sugi and Shin, Jaemin and Kim, Yusung},
  journal={IEEE Internet of Things Journal}, 
  title={Self-Attention-based Uplink Radio Resource Prediction in 5G Dual Connectivity}, 
  year={2023},
  volume={},
  number={},
  pages={1-1},
  doi={10.1109/JIOT.2023.3283490}}

@inproceedings{10.1145/3386901.3388911,
author = {Lee, Jinsung and Lee, Sungyong and Lee, Jongyun and Sathyanarayana, Sandesh Dhawaskar and Lim, Hyoyoung and Lee, Jihoon and Zhu, Xiaoqing and Ramakrishnan, Sangeeta and Grunwald, Dirk and Lee, Kyunghan and Ha, Sangtae},
title = {PERCEIVE: Deep Learning-Based Cellular Uplink Prediction Using Real-Time Scheduling Patterns},
year = {2020},
isbn = {9781450379540},
publisher = {Association for Computing Machinery},
address = {New York, NY, USA},
url = {https://dl.acm.org/10.1145/3386901.3388911},
doi = {10.1145/3386901.3388911},
booktitle = {Proceedings of the 18th International Conference on Mobile Systems, Applications, and Services},
pages = {377–390},
numpages = {14},
location = {Toronto, Ontario, Canada},
series = {MobiSys '20}
}

@ARTICLE{8051088,
  author={Yue, Chaoqun and Jin, Ruofan and Suh, Kyoungwon and Qin, Yanyuan and Wang, Bing and Wei, Wei},
  journal={IEEE Transactions on Mobile Computing}, 
  title={LinkForecast: Cellular Link Bandwidth Prediction in LTE Networks}, 
  year={2018},
  volume={17},
  number={7},
  pages={1582-1594},
  doi={10.1109/TMC.2017.2756937}}

@misc{zte_corporation_2023, title={AIS and ZTE announce Thailand’s first 5G mmWave SA showcase at 26GHz}, url={https://www.zte.com.cn/global/about/news/ais-and-zte-announce-thailands-first-5g-mmwave-sa-showcase-at-26ghz.html}, journal={www.zte.com.cn}, author={ZTE Corporation}, year={2023}, month={Jun} }

@misc{qualcomm_2023, title={Snapdragon X75 5G Modem-RF System}, url={https://www.qualcomm.com/products/technology/modems/snapdragon-x75-5g-modem-rf-system}, journal={www.qualcomm.com}, author={Qualcomm}, year={2023} }

@misc{qualcomm_x65, title={Snapdragon X65 5G Modem-RF System}, url={https://www.qualcomm.com/products/technology/modems/snapdragon-x65-5g-modem-rf-system}, journal={www.qualcomm.com}, author={Qualcomm}, year={2021} }

@INPROCEEDINGS{10118777,
  author={Arunruangsirilert, Kasidis and Wongprasert, Pasapong and Katto, Jiro},
  booktitle={2023 IEEE Wireless Communications and Networking Conference (WCNC)}, 
  title={Performance Evaluations of C-Band 5G NR FR1 (Sub-6 GHz) Uplink MIMO on Urban Train}, 
  year={2023},
  volume={},
  number={},
  pages={1-6},
  doi={10.1109/WCNC55385.2023.10118777}}

@standard{itu_m2083,
address = {Geneva, Switzerland},
type = {Standard},
title = {Minimum requirements related to technical performance for IMT-2020 radio interface(s)},
shorttitle = {{ISO}/{IEC} {TR} 29110-1:2016},
url = {https://www.itu.int/pub/R-REP-M.2410},
language = {en},
number = {Rec. ITU-R M.2083},
institution = {International Telecommunication Union},
year = {Nov. 2017}
}

@inproceedings{10.5555/2969239.2969329,
author = {Shi, Xingjian and Chen, Zhourong and Wang, Hao and Yeung, Dit-Yan and Wong, Wai-kin and Woo, Wang-chun},
title = {Convolutional LSTM Network: A Machine Learning Approach for Precipitation Nowcasting},
year = {2015},
publisher = {MIT Press},
address = {Cambridge, MA, USA},
booktitle = {Proceedings of the 28th International Conference on Neural Information Processing Systems - Volume 1},
pages = {802–810},
numpages = {9},
location = {Montreal, Canada},
series = {NIPS'15}
}

@misc{ray_2020, title={T‑Mobile Is Still the Only Adult in the Room on 5G. And Here’s Why. ‑ T‑Mobile Newsroom}, url={https://www.t-mobile.com/news/network/t-mobile-5g-coverage-and-speed}, journal={T-Mobile Newsroom}, author={Ray, Neville}, year={2020}, month={Nov} }

@standard{3GPP_38-101-1,
type = {Standard},
title = {NR; User Equipment (UE) radio transmission and reception; Part 1: Range 1 Standalone},
language = {en},
number = {TS 38.101-1 version 16.5.0 Release 16},
institution = {3GPP},
year = {Nov. 2022}
}

@INPROCEEDINGS{9488851,
  author={Sacco, Alessio and Flocco, Matteo and Esposito, Flavio and Marchetto, Guido},
  booktitle={IEEE INFOCOM 2021 - IEEE Conference on Computer Communications}, 
  title={Owl: Congestion Control with Partially Invisible Networks via Reinforcement Learning}, 
  year={2021},
  volume={},
  number={},
  pages={1-10},
  doi={10.1109/INFOCOM42981.2021.9488851}}

@inproceedings{10.1145/2910017.2910608,
author = {Kurdoglu, Eymen and Liu, Yong and Wang, Yao and Shi, Yongfang and Gu, ChenChen and Lyu, Jing},
title = {Real-Time Bandwidth Prediction and Rate Adaptation for Video Calls over Cellular Networks},
year = {2016},
isbn = {9781450342971},
publisher = {Association for Computing Machinery},
address = {New York, NY, USA},
url = {https://doi.org/10.1145/2910017.2910608},
doi = {10.1145/2910017.2910608},
booktitle = {Proceedings of the 7th International Conference on Multimedia Systems},
articleno = {12},
numpages = {11},
location = {Klagenfurt, Austria},
series = {MMSys '16}
}

@inproceedings{10.1145/3300061.3345430,
author = {Zhou, Anfu and Zhang, Huanhuan and Su, Guangyuan and Wu, Leilei and Ma, Ruoxuan and Meng, Zhen and Zhang, Xinyu and Xie, Xiufeng and Ma, Huadong and Chen, Xiaojiang},
title = {Learning to Coordinate Video Codec with Transport Protocol for Mobile Video Telephony},
year = {2019},
isbn = {9781450361699},
publisher = {Association for Computing Machinery},
address = {New York, NY, USA},
url = {https://doi.org/10.1145/3300061.3345430},
doi = {10.1145/3300061.3345430},
booktitle = {The 25th Annual International Conference on Mobile Computing and Networking},
articleno = {29},
numpages = {16},
location = {Los Cabos, Mexico},
series = {MobiCom '19}
}

\end{document}